\documentclass[a4paper,fleqn]{cas-sc}

\usepackage[numbers]{natbib}

\def\tsc#1{\csdef{#1}{\textsc{\lowercase{#1}}\xspace}}
\tsc{WGM}
\tsc{QE}

\usepackage{amssymb}
\usepackage{amsmath}
\usepackage{latexsym}
\usepackage{amsthm}
\usepackage{epigraph}
\usepackage{booktabs}
\usepackage{braket}
\usepackage{appendix}
\usepackage{wrapfig}
\usepackage{graphicx}
\usepackage{pdflscape}
\usepackage{epstopdf}
\usepackage{pdfpages}
\usepackage{pgfplots}
\usepackage{eurosym}
\usepackage{array,multirow}
\usepackage{url}

\usepackage[font={footnotesize}]{caption}
\usepackage[font={scriptsize}]{subcaption}
\usepackage{comment}

\newcommand{\norm}[1]{\left\lVert#1\right\rVert}
\DeclareMathOperator*{\argmax}{arg\,max}
\newcommand{\abs}[1]{\left\lvert#1\right\rvert}

\usepackage{soul}
\soulregister\cite7
\soulregister\ref7
\soulregister\eqref7
\soulregister\pageref7
\usepackage{cancel}
\usepackage{bbm}

\begin{document}
\let\WriteBookmarks\relax
\def\floatpagepagefraction{1}
\def\textpagefraction{.001}

\shorttitle{A Robust SVM for Raman COVID-19 Data}    

\shortauthors{Piazza M., Spinelli A., Maggioni F., Bedoni M. and Messina E.}  

\title [mode = title]{A Robust Support Vector Machine Approach for Raman Data Classification}  



%

\author[1]{Marco Piazza}
\author[2]{Andrea Spinelli}
\author[2]{Francesca Maggioni}
\author[3]{Marzia Bedoni}
\author[1]{Enza Messina}
\ead{enza.messina@unimib.it}
\cormark[1]

\affiliation[1]{organization={Department of Informatics, Systems and Communication, University of Milano-Bicocca},
            addressline={Viale Sarca 336}, 
            city={Milan},
            citysep={},
            postcode={20126}, 
            state={Italy}}
\affiliation[2]{organization={Department of Management, Information and Production Engineering, University of Bergamo},
            addressline={Viale G. Marconi 5}, 
            city={Dalmine},
            citysep={},
            postcode={24044}, 
            state={Italy}}


\affiliation[3]{organization={IRCCS Fondazione Don Carlo Gnocchi ONLUS},
            addressline={Via Capecelatro 66}, 
            city={Milan},
            citysep={},
            postcode={20148}, 
            state={Italy}}




\begin{abstract}
Recent advances in healthcare technologies have led to the availability of large amounts of biological samples across several techniques and applications. In particular, in the last few years, Raman spectroscopy analysis of biological samples has been successfully applied for early-stage diagnosis. However, spectra' inherent complexity and variability make the manual analysis challenging, even for domain experts. For the same reason, the use of traditional Statistical and \textit{Machine Learning} (ML) techniques could not guarantee for accurate and reliable results. ML models, combined with robust optimization techniques, offer the possibility to improve the classification accuracy and enhance the resilience of predictive models. In this paper, we investigate the performance of a novel robust formulation for Support Vector Machine (SVM) in classifying COVID-19 samples obtained from Raman Spectroscopy. Given the noisy and perturbed nature of biological samples, we protect the classification process against uncertainty through the application of robust optimization techniques. Specifically, we derive robust counterpart models of deterministic formulations using bounded-by-norm uncertainty sets around each observation. We explore the cases of both linear and kernel-induced classifiers to address binary and multiclass classification tasks. The effectiveness of our approach is validated on real-world COVID-19 datasets provided by Italian hospitals by comparing the results of our simulations with a state-of-the-art classifier.
\end{abstract}


\maketitle

\section{Introduction}

Raman Spectroscopy (RS) is a technique based on the inelastic scattering of monochromatic light to observe low-frequency modes in a molecular system (see \cite{ButAshBirCinCurDorEsmFullNigGarMar2016}). The resulting scattering pattern serves as a ``fingerprint", revealing information about the sample's chemical composition, including the presence, concentrations, and interactions of its molecules. In the healthcare field, the spectral information acquired from biological samples can be exploited to diagnose and monitor the emergence of pathologies by detecting certain biomarkers associated with the suspected condition. These applications typically involve a variety of target samples, such as blood-based fluids (serum, plasma) or human tissues (see \cite{covid_one,Chen2019,Zhang2021}). In recent research (see, for instance, \cite{als_paper, pd_paper}), the analysis of saliva samples has demonstrated potential for identifying the presence of relevant biomarkers and their concentration, making saliva one of the most promising targets for analysis, especially considering its straightforward accessibility.

Labelled spectra coming from RS analysis have been successfully used to train Machine Learning or \textit{Deep Learning} (DL) models.  
Despite their potential, Deep Learning methods require vast amounts of data for effective training. This represents a significant challenge in the field of Spectroscopy, where the acquisition of data samples may require considerable financial, human, and time resources, potentially compromising the applicability of these models.
On the other hand among the plethora of ML algorithms that have been designed to handle classification problems, Support Vector Machines (SVMs) have received strong attention (see \cite{PimOspAra2024}).
SVM models have been extensively employed to handle classification tasks in medicine and healthcare applications (see \cite{AkiOlaAriFasMbuOkpOwa2023,GuiFerLofCon2024,TrkumBhaKhaAlkVicVer2024}). In particular, in Raman Spectroscopy context, SVM algorithms have recently been investigated with the aim of providing fast and efficient solutions for the early diagnosis of various diseases (see \cite{CheWuCheLuoShiLiLvCheSuWu2023,DucTungPhanNgu2023}).
Introduced in \cite{VapChe1974}, SVM aims to find the best separating hyperplane that maximizes the margin between classes. To improve the classification accuracy, many SVM-based models have been proposed in the literature (see, for instance, \cite{Man1998,JayKheCha2007,Peng2011,Jim-CorMorPin2021}). In this paper, we focus on the variant introduced in \cite{LiuPot2009} and further extended in \cite{maggspin}. The strength of this approach over other SVMs lies in its two-step procedure. Indeed, rather than constructing a single hyperplane, the method first separates the data using two parallel hyperplanes derived as solutions of a SVM model. The optimal final hyperplane is then searched within the region between these two, minimizing the total number of misclassified data points. 

The interpretation of spectra obtained from RS, particularly salivary samples, can be challenging due to the complex combination of several basic molecules, which results in a high sensitivity to noise and a possible low signal-to-noise ratio \cite{raman_preproc}. A number of preprocessing steps have been proposed in the literature \cite{raman_preproc} to address this challenge. While some noise sources, such as outliers and spikes, have been effectively addressed, the resulting data remains significantly noisy.
For this reason, it is crucial to deal with ML models able to protect the classification process against such perturbations. In the mathematical programming literature, various techniques have been developed to address the problem of uncertainty in ML methods. Among these, \textit{Robust Optimization} (RO) is widely recognized as one of the main paradigms (see \cite{Ben-TalElGNem2009,BerBroCar2010}). RO assumes that all potential realizations of the uncertain parameters fall within a predefined uncertainty set. The corresponding robust model is then derived by optimizing against the worst-case realizations of the parameters across the entire uncertainty set (see \cite{BerDunPawZhu2019}). The application of RO techniques generally leads to improved predictive performance of the ML methods (see \cite{MalLopVai2020,FacMagPot2022}).

To this aim, in this work, we investigate the performance of a novel robust formulation for SVM introduced in \cite{maggspin} to facilitate the classification of Raman spectra. We conduct a comparative analysis between the proposed method and the classical SVM approach on a classification task aimed at diagnosing COVID-19 from real-world saliva samples. The computational experiments demonstrate that the robust SVM model exhibits superior performance compared to the baseline in the majority of investigated conditions, making it a suitable candidate for Raman Spectroscopy analysis under uncertainty.

The remainder of the paper is organized as follows. Section \ref{sec_related_works} reviews the existing literature on the problem. In Section \ref{sec_mathematical_models}, the mathematical models and their robust counterpart are presented. Section \ref{sec_computational_results} describes data collection and reports the experimental study. Finally, Section \ref{sec_conclusions} concludes the paper and discusses future works.

\section{Related Works} \label{sec_related_works}

Currently, the automatic classification of spectral data is predominantly performed using ML models, with a considerable proportion of these being linear models. 
Considering the high dimensionality of spectral data often these methods are combined with feature reduction strategies, such as \textit{Principal Component Analysis} (PCA, see \cite{Hotelling1933}). In recent years, some works also explored the possibility of classifying Raman Spectra with DL algorithms (see \cite{Lussier2020}). However, the collection of spectral datasets is a time-consuming and costly process, and given the considerable data requirements for the effective training of a Deep Neural Network, the use of ML models remains the predominant approach. 

According to the recent literature (see \cite{Lussier2020}), three main fields of application are identified as the most common in combining ML and spectroscopy. The first is the food industry, where ML is used to detect fraud and identify product alterations (see \cite{56}). The second is forensic science, where ML is employed to identify illicit drugs (see \cite{66}) or analyse criminal scenes (see \cite{70}). The third is medicine and healthcare, where ML is used to recognize bacteria and viruses or provide automatic diagnosis. Given the focus of this work, the remainder of this section will concentrate on the domain of healthcare, with a particular emphasis on the development of automated techniques for the diseases' diagnosis. 

In the field of healthcare, the predominant approach to classification is through the use of linear models, particularly \textit{Linear Discriminant Analysis} (LDA, see \cite{Carlomagno_copd, Diao2023}) and SVM (see \cite{Zheng2018, Rahman2022, He2021}). There is a limited number of studies that have employed unsupervised algorithms, such as clustering (see \cite{Sbroscia2020}) and $k$-Nearest Neighbors (see \cite{Cui2018}), and only recently have some studies attempted to implement neural networks (see \cite{cmess, pd_paper}). Often ML approaches are combined with feature reduction to allow them to deal with the highly dimensionality and noise characterizing spectral data. In this sense, the choice is almost completely for PCA. 
One of the most interesting applications of RS and ML is cancer detection and diagnosis. In recent years, cancers including liver (see \cite{90}) or thyroid (see \cite{Zheng2018}) have been investigated through the use of Raman data and multivariate analysis. Other examples includes the diagnosis of bladder cancer with SVM (see \cite{Chen2019}) or DL models (see \cite{Kazemzadeh2022}). Breast cancer is often investigated with various automatic algorithms, such as LDA (see \cite{Talari2019}), SVM (see \cite{Zhang2022}) or DL models (see \cite{Kothari2021}). Examples of other applications are neurodegenerative diseases, such as Alzheimer and Parkinson, with SVM and tree-based ensemble (see \cite{pd_paper}), and Amyotrophic Lateral Sclerosis (see \cite{als_paper}). 
Following the global diffusion of the novel coronavirus, numerous studies have investigated the potential for automated identification of infected individuals, leveraging the integration of RS extracted from diverse biological samples, including saliva and blood, in conjunction with ML algorithms. Two illustrative examples include \cite{covid_one}, where a \textit{Light Gradient Boosting Machine} is trained to recognize spectra from blood serum and \cite{Karunakaran_2022}, where a SVM is employed to classify spectra extracted from saliva.

All the ML approaches discussed so far implicitly assume that input data are precisely known at the moment of classifying. However, this assumption is often unrealistic with real-world observations, especially when dealing with spectra coming from RS or saliva samples. Indeed, such data are frequently plagued by noise and perturbations, resulting in worsening performances of the classification process. To address the problem of uncertainty in training samples, RO techniques have been developed in the ML literature to help preventing the worsening of the solution quality (see \cite{XuCarMan2009}). Robust formulations of standard classification methods including logistic regression, SVM and decision trees are discussed in \cite{BerDunPawZhu2019}. RO techniques have been also applied to other variants of the classical SVM model. The robust counterpart of the linear approach presented in \cite{LiuPot2009} is extended in \cite{FacMagPot2022}, introducing a novel \emph{Distributionally Robust Optimization} (DRO) formulation with moment-based ambiguity sets. An application of such robust and distributionally robust SVM methodology for COVID-19 patient classification is presented in \cite{MagFacGheManBonORAHS2022}. In \cite{maggspin} the robust extension of the approach designed in \cite{LiuPot2009} is developed employing kernel-induced decision boundaries and bounded-by-$\ell_p$-norm uncertainty sets. The performance of the approach is tested on a real-world vehicle emissions task (see \cite{MagSpiODS2024}). In \cite{PengXu2013} a robust version of the \emph{TWin Support Vector Machine} (TWSVM, \cite{JayKheCha2007}) classifier is proposed, incorporating uncertainty in the variance matrices of the two classes. The robust extension of TWSVM, formulated as a \textit{Second Order Cone Programming} problem (SOCP), is presented in \cite{QiTianShi2013}. Additionally, the recent work of \cite{DelMagSpi2024TPMSVM} introduces a robust and multiclass extension of the \textit{Twin Parametric Margin Support Vector Machine} (TPMSVM, see \cite{Peng2011}), with an application in the field of sustainability (see \cite{DeLMagSpiLOD2024}). Finally, \cite{Kha-ShiBab-AzaHos-NodPar2023} and \cite{LinFangFangGao2024,LinFangFangGaoLuo2024} explore the integration of \emph{Chance-Constrained Programming} (CCP) and DRO techniques into linear and nonlinear SVM models, respectively, accounting for uncertain data.

Past research has investigated the potential of combining ML methods, particularly classical SVM approaches, with spectral data for diagnostic purposes. However, as far as we know, none of these studies have tackled the problem of considering perturbations and noise commonly found in saliva samples, which pose challenges for traditional ML models in processing this type of data. In this work, we assess the performance of the robust SVM approach proposed in \cite{maggspin} in handling saliva Raman spectra under data uncertainty. To the best of our knowledge, this is the first contribution in the literature that directly incorporates uncertainty into a SVM model using real data coming from RS analysis. 

\section{Mathematical Models} \label{sec_mathematical_models}
In this section, we describe the Support Vector Machine (SVM) models proposed in \cite{maggspin} and based on the works in \cite{Man1998,LiuPot2009} for addressing classification problems with nonlinear decision boundaries. We start by examining the deterministic formulations for binary and multiclass classification tasks (Section \ref{sec_deterministic_SVM}). Next, we consider the robust counterpart models in the context of bounded-by-$\ell_p$-norm uncertainty sets (Section \ref{sec_robust_SVM}).

\subsection{Deterministic Formulation} \label{sec_deterministic_SVM}
Let $\{(x^{(i)},y^{(i)})\}_{i=1}^m$ be the set of training data points, where $x^{(i)}\in \mathbb{R}^n$ is the vector of features, and $y^{(i)}\in\{-1,1\}$ is the label representing the class to which the $i$-th data point belongs.

The aim of the model proposed in \cite{Man1998} is to find the best separating hypersurface as solution of the following $\ell_1$-SVM formulation:
\begin{equation} \label{model_SVM_deterministic_binary}
\begin{aligned}
\min_{u,\gamma,\xi}  \quad & \norm{u}_1+\nu \sum_{i=1}^m\xi_i\\
\text{s.t.} \quad & y^{(i)}\bigg(\sum_{j=1}^m K_{ij}y^{(j)}u_j-\gamma\bigg)\geq 1-\xi_i & \quad  i=1,\ldots,m\\
 & \xi_i\geq 0 & \quad  i=1,\ldots,m,
\end{aligned}
\end{equation}
where $\nu$ is a positive parameter balancing the terms in the objective function, $K_{ij}:=k(x^{(i)},x^{(j)})$ is the Gram matrix defined according to kernel function $k:\mathbb{R}^n\times\mathbb{R}^n\to \mathbb{R}$ (see Table \ref{tab_kernel}), and $\xi\in\mathbb{R}^m$ is a slack vector. The kernel function $k(\cdot,\cdot)$ is associated with a feature map $\phi:\mathbb{R}^n \to \mathcal{H}$ that projects data from the \emph{input space} $\mathbb{R}^n$ to a higher-dimensional space $\mathcal{H}$, called \emph{feature space}, and equipped with the norm $\norm{\cdot}_{\mathcal{H}}$. For a comprehensive overview on kernel functions applied to ML methods, the reader is referred to \cite{SchSmo2001}.

\begin{table}[h!]
    \centering
    \begin{tabular}{c|c|c}
    \toprule
        Kernel function & $k(x,x')$ & Parameters\\ \hline
        Homogeneous polynomial & $\langle x,x' \rangle^d$ & $d \in \mathbb{N}$ \\ \hline
        Inhomogeneous polynomial & $(\langle x,x' \rangle+c)^d$ & $c \in\mathbb{R}^+,d \in \mathbb{N}$\\ \hline
        Gaussian & $\exp(-\frac{\|x-x'\|_2^2}{2\alpha^2})$ & $\alpha \in \mathbb{R}_0^+$\\ \bottomrule
    \end{tabular}
    \caption{Examples of kernel functions typically used to train SVM models.}
    \label{tab_kernel}
\end{table}

Once $u,\gamma,\xi$ are obtained as solutions of \eqref{model_SVM_deterministic_binary}, an initial nonlinear decision boundary $S_0:=(u,\gamma)$ is defined according to the following equation:
\begin{equation}\label{nonlinear_hyperplane}
\sum_{i=1}^m k(x,x^{(i)})y^{(i)}u_i=\gamma.
\end{equation}

Similarly to \cite{LiuPot2009}, for each class the greatest misclassification error is computed through formulas:
\begin{equation} \label{omega_nonlinear}
\omega_{1}:= \max_{i=1,\ldots,m} {(D\xi)}_i \qquad \omega_{-1}:= \max_{i=1,\ldots,m} {(-D\xi)}_i,
\end{equation}

where $D$ is a diagonal matrix with entries $D_{ii}:=y^{(i)}$, for all $i=1,\ldots,m$. The values $\omega_{1}$ and $\omega_{-1}$ are used to shift the initial hypersurface $S_0$, leading to $S_{1}:=(u, \gamma - 1+\omega_1)$ and $S_{-1}:=(u,\gamma+1-\omega_{-1})$ defined as follows:
\begin{equation}
    \label{S1_S_minus1}
    \begin{aligned}    
        S_{1}: & \sum_{i=1}^{m} k(x,x^{(i)})y^{(i)}u_i=\gamma-1+\omega_1 \\
        S_{-1}: & \sum_{i=1}^{m} k(x,x^{(i)})y^{(i)}u_i=\gamma+1-\omega_{-1}.
    \end{aligned}
\end{equation}

Finally, the optimal kernel-induced decision boundary $S:=(u,b)$ lies in the region between $S_{1}$ and $S_{-1}$, being $b$ the solution of the following problem:
\begin{equation}\label{linesearch_det_binary}
\begin{aligned}
\min_{b}  \quad & \sum_{i=1}^m \mathbbm{1}\bigg(y^{(i)}b-y^{(i)}\sum_{j=1}^m K_{ij}y^{(j)}u_j\bigg)\\
\text{s.t.} \quad & \gamma+1-\omega_{-1}\leq b \leq \gamma-1+\omega_{1},
\end{aligned}
\end{equation}
where $\mathbbm{1}(\cdot):\mathbb{R}\to \{0,1\}$ is the indicator function. From a computational standpoint, the solution of model \eqref{linesearch_det_binary} is obtained via a linear search procedure. In particular, the interval $[\gamma+1-\omega_{-1},\gamma-1+\omega_{1}]$ is partitioned into $N_{\max}$ equally spaced sub-intervals, and the objective function is evaluated on each of them. The optimal solution $b$ corresponds to the one yielding the minimum value of the objective function across all sub-intervals. Finally, every new observation $x\in\mathbb{R}^n$ is classified according to the decision function $\mathbbm{1}\big(\sum_{i=1}^m k(x,x^{(i)})y^{(i)}u_i-b\big)$.

\begin{figure}[h!]
     \centering
  \begin{subfigure}[b]{0.49\textwidth}
         \centering
         \includegraphics[width=\textwidth]{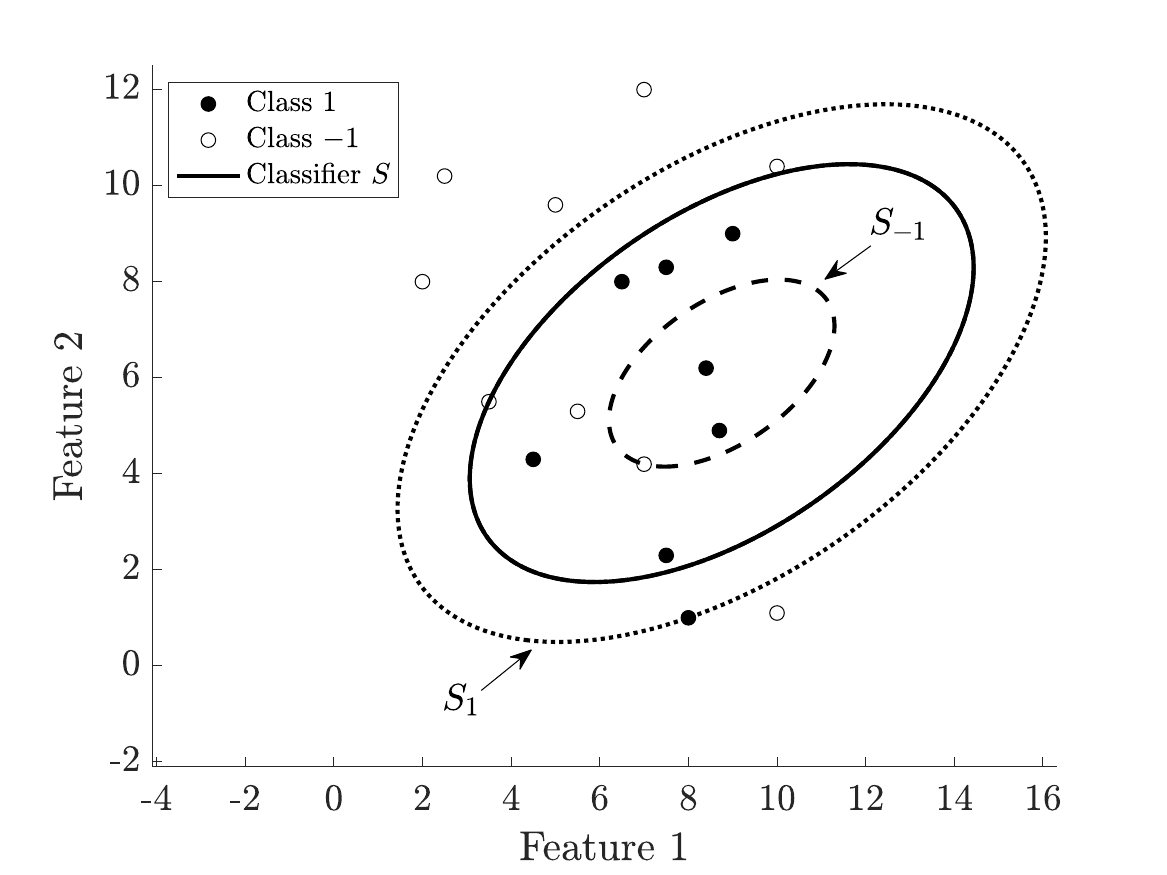}
         \caption{Binary classification with inhomogeneous quadratic kernel.}
         \label{fig_toy_2d_det_binary}
     \end{subfigure}
     \begin{subfigure}[b]{0.49\textwidth}
         \centering
\includegraphics[width=\textwidth]{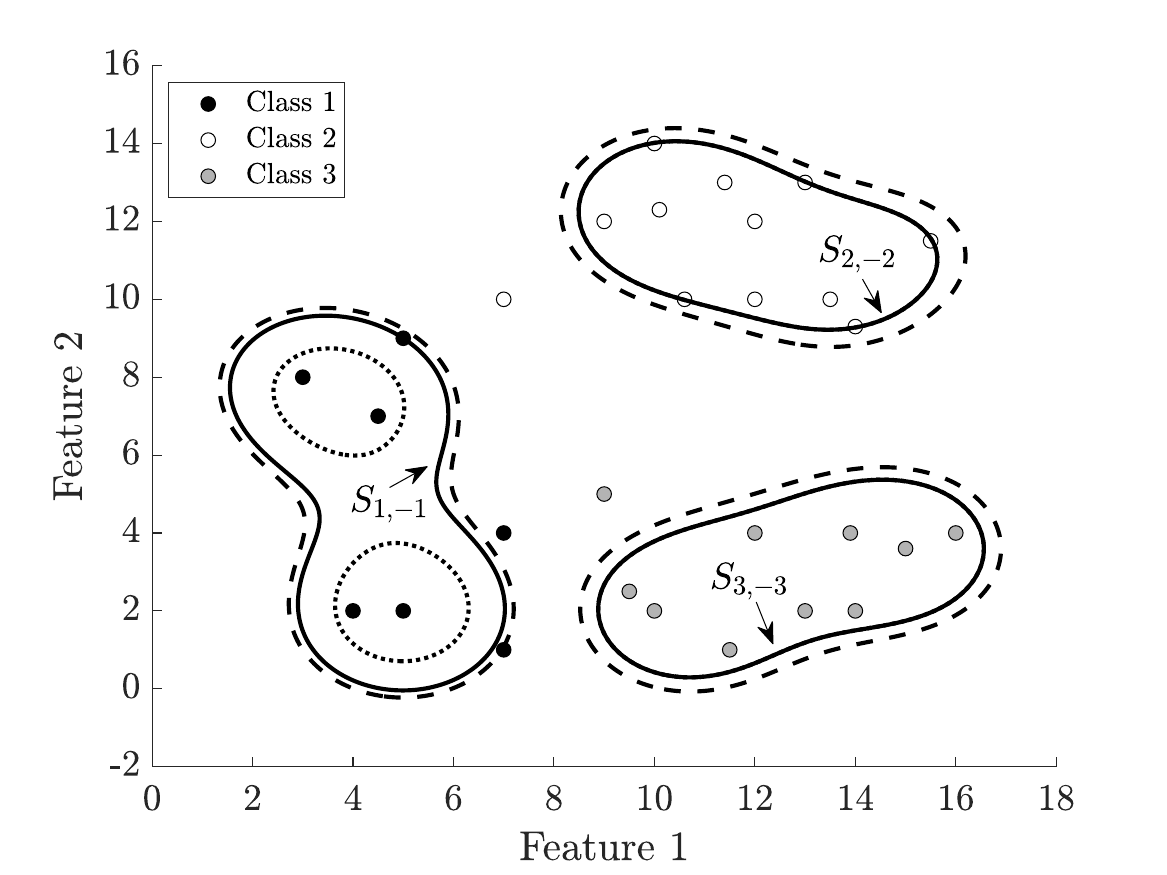}
         \caption{Multiclass classification with Gaussian kernel.}
         \label{fig_toy_2d_det_multiclass}
     \end{subfigure}
          \caption{Separating surfaces obtained with inhomogeneous quadratic kernel ($d=2$, $c=0.3$) for binary classification (on the left) and Gaussian kernel ($\alpha=1.9$) for multiclass classification (on the right). For each class $l=1,2,3$, the dotted line and the dashed line represent respectively $S_{l}$ and $S_{-l}.$}
        \label{fig_toy_2d_det}
\end{figure}

As an example, in Figure \ref{fig_toy_2d_det_binary}, we show the separating surfaces resulting from the application of the considered SVM methodology to a two-dimensional toy problem. In model \eqref{model_SVM_deterministic_binary}, we set $\nu=1$, and consider inhomogeneous quadratic kernel, with $d=2$ and $c=0.3$ (see Table \ref{tab_kernel}).

In the case of multiclass classification tasks, a \emph{one-versus-all} approach is considered, classifying training data points of each class against all the other classes. Formally, let $y^{(i)}\in\{1,\ldots,L\}$ be the label of the $i$-th observations, with $L$ the number of classes. For each class $l=1,\ldots,L$, an initial separating hypersurface $S_{l,0}:=(u_l,\gamma_l)$ is constructed as in \eqref{nonlinear_hyperplane}, where $u_l\in\mathbb{R}^n$ and $\gamma_l\in\mathbb{R}$ are the solutions of the following multiclass version of model \eqref{model_SVM_deterministic_binary}:
\begin{equation} \label{model_SVM_deterministic_multiclass}
\begin{aligned}
\min_{u_l,\gamma_l,\xi_l}  \quad & \norm{u_l}_1+\nu \sum_{i=1}^m\xi_{l,i}\\
\text{s.t.} \quad & \widehat{y}^{(i)}_l\bigg(\sum_{j=1}^m K_{ij}\widehat{y}^{(j)}_lu_{l,j}-\gamma_l\bigg)\geq 1-\xi_{l,i} & \quad  i=1,\ldots,m\\
 & \xi_{l,i}\geq 0 & \quad  i=1,\ldots,m,
\end{aligned}
\end{equation}
with $\widehat{y}^{(i)}_l=1$ if $y^{(i)}=l$, and $\widehat{y}^{(i)}_l=-1$ otherwise. Then, the diagonal matrix $\widehat{D}_l$, with $\widehat{D}_{l,ii}:=\widehat{y}^{(i)}_l$, $i=1,\ldots,m$, is constructed and the multiclass equivalent formulas of \eqref{omega_nonlinear} are computed as follows:
\begin{equation*}
\omega_{l}:= \max_{i=1,\ldots,m} {(\widehat{D}_l\xi_l)}_i \qquad \omega_{-l}:= \max_{i=1,\ldots,m} {(-\widehat{D}_l\xi_l)}_i.
\end{equation*}

Hypersurface $S_{l,0}$ is then shifted to get $S_{l}:=(u_l,\gamma_l-1+\omega_l)$ and $S_{-l}:=(u_l,\gamma_l+1-\omega_{-l})$ (see equation \eqref{S1_S_minus1}). Finally, the optimal decision boundary for class $l$ versus all the others is $S_{l,-l}:=(u_l,b_l)$, being $b_l$ the solution of the following model:
\begin{equation}\label{linesearch_det_multiclass}
\begin{aligned}
\min_{b_l}  \quad & \sum_{i=1}^m \mathbbm{1}\bigg(\widehat{y}^{(i)}_lb_l-\widehat{y}^{(i)}_l\sum_{j=1}^m K_{ij}\widehat{y}^{(j)}_lu_{l,j}\bigg)\\
\text{s.t.} \quad & \gamma_l+1-\omega_{-l}\leq b_l \leq \gamma_l-1+\omega_{l}.
\end{aligned}
\end{equation}

The decision function of the $l$-th class, with $l=1,\ldots,L$, is given by $f_l(x):=\sum_{i=1}^m k(x,x^{(i)})\widehat{y}^{(i)}_lu_{l,i}-b_l$, and each new observation $x\in\mathbb{R}^n$ is assigned to the class $l^*:=\argmax_{l= 1,\ldots,L} f_l(x)$ (see \cite{LopMalCar2017}).

We represent in Figure \ref{fig_toy_2d_det_multiclass} the results of the approach for a multiclass classification problem involving three distinct classes. We consider $\nu=1$ in model \eqref{model_SVM_deterministic_multiclass} and Gaussian kernel with parameter $\alpha=1.9$ (see Table \ref{tab_kernel}).

\subsection{Robust Formulation} \label{sec_robust_SVM}

In this section, we discuss the robust counterpart of the deterministic approaches discussed so far and derived in \cite{maggspin}. According to the robust optimization framework, we assume that input data are plagued by unknown perturbations and construct an uncertainty set around each observation. The best solution is the one optimizing against the worst-case realization across the entire uncertainty set (see \cite{BerDunPawZhu2019}).

Formally, let each observation $x^{(i)}$ in the input space $\mathbb{R}^n$ be subject to an additive and unknown perturbation vector $\sigma^{(i)}$, whose $\ell_p$-norm, with $p\in[1,\infty]$, is bounded by a nonnegative constant $\eta^{(i)}$. As a result, the uncertainty set around $x^{(i)}$ can be written as follows:
\begin{equation} \label{U_input_space}
\mathcal{U}_p(x^{(i)}):=\Set{ x\in\mathbb{R}^n : x=x^{(i)}+\sigma^{(i)}, \lVert \sigma^{(i)} \lVert_p \leq \eta^{(i)} }.
\end{equation}

The parameter $\eta^{(i)}$ regulates the degree of conservatism: if $\eta^{(i)}=0$, then $\sigma^{(i)}$ is the zero vector of $\mathbb{R}^n$ and $\mathcal{U}_p(x^{(i)})$ coincides with $x^{(i)}$. Common choices for the $\ell_p$-norm in the robust optimization literature include $p=1,2,\infty$, leading to polyhedral, spherical and box uncertainty sets, respectively.

To extend this construction to the feature space $\mathcal{H}$, we assume that the uncertainty set around the projected data $\phi(x^{(i)})$ is modelled as:
\begin{equation} \label{U_feature_space}
\mathcal{U}_{\mathcal{H}}\big(\phi(x^{(i)})\big):=\Set{ z\in\mathcal{H} : z=\phi(x^{(i)})+\zeta^{(i)}, \lVert \zeta^{(i)} \lVert_{\mathcal{H}} \leq \delta^{(i)} },
\end{equation}
where the perturbation $\zeta^{(i)}$ belongs to $\mathcal{H}$ and its $\mathcal{H}$-norm is bounded a nonnegative constant $\delta^{(i)}$. The latter may be unknown but it depends on the known bound $\eta^{(i)}$ in the input space: if no uncertainty occurs in the input space ($\eta^{(i)}=0$), no uncertainty will occur in the feature space too ($\delta^{(i)}=0$). The relation between $\delta^{(i)}$ and $\eta^{(i)}$ has been explored in \cite{maggspin} where a closed-form expression of $\delta^{(i)}$ has been derived as function of $\eta^{(i)}$ for typically used kernel functions.

Once the uncertainty sets \eqref{U_input_space}-\eqref{U_feature_space} have been constructed, it is possible to derive the robust counterparts of models \eqref{model_SVM_deterministic_binary} and \eqref{model_SVM_deterministic_multiclass}. Specifically, for binary classification tasks, the robust model is given by:
\begin{equation} \label{model_SVM_robust_binary}
\begin{aligned}
\min_{u,\gamma,\xi}  \quad & \norm{u}_{1}+\nu \sum_{i=1}^m\xi_i\\
\text{s.t.} \quad & y^{(i)}\sum_{j=1}^m K_{ij}y^{(j)}u_j -\delta^{(i)}\sum_{j=1}^{m} \sqrt{K_{jj}} \abs{u_j} \geq 1-\xi_i+y^{(i)}\gamma & i=1,\ldots,m\\
 & \xi_i\geq 0 &  i=1,\ldots,m.
\end{aligned}
\end{equation}

Similar to the deterministic framework, once $u$, $\gamma$ and $\xi$ are determined as solutions of \eqref{model_SVM_robust_binary}, then $\omega_{1}$ and $\omega_{-1}$ are calculated using the expressions in \eqref{omega_nonlinear}. Finally, the optimal separating hypersurface $S=(u,b)$ is obtained, being $b$ the optimal solution of the following robust version of problem $\eqref{linesearch_det_binary}$:
\begin{equation} \label{linesearch_rob_binary}
\begin{aligned}
\min_{b}  \quad & \sum_{i=1}^m \mathbbm{1}\bigg[\bigg(y^{(i)}b-y^{(i)}\sum_{j=1}^m K_{ij}y^{(j)}u_j+\delta^{(i)}\sum_{j=1}^{m} \sqrt{K_{jj}} \abs{u_j}\bigg)_i\bigg]\\
\text{s.t.} \quad & \gamma+1-\omega_{-1}\leq b \leq \gamma-1+\omega_{1}.
\end{aligned}
\end{equation}

When addressing a multiclass classification problem, the robust extension of model \eqref{model_SVM_deterministic_multiclass} for the $l$-th class, with $l=1,\ldots,L$, is expressed as follows:
\begin{equation} \label{model_SVM_robust_multiclass}
\begin{aligned}
\min_{u_l,\gamma_l,\xi_l}  \quad & \norm{u_l}_{1}+\nu \sum_{i=1}^m\xi_{l,i}\\
\text{s.t.} \quad & \widehat{y}^{(i)}_l\sum_{j=1}^m K_{ij}\widehat{y}^{(j)}_lu_{l,j} -\delta^{(i)}\sum_{j=1}^{m} \sqrt{K_{jj}} \abs{u_{l,j}} \geq 1-\xi_{l,i}+\widehat{y}^{(i)}_l\gamma_l & \quad i=1,\ldots,m\\
 & \xi_{l,i}\geq 0 & \quad  i=1,\ldots,m.
\end{aligned}
\end{equation}

The optimal parameter $b_l$ of the kernel-induced decision boundary $S_{l,-l}:=(u_l,b_l)$ is the solution of:
\begin{equation} \label{linesearch_rob_multiclass}
\begin{aligned}
\min_{b_l}  \quad & \sum_{i=1}^m \mathbbm{1}\bigg[\bigg(\widehat{y}^{(i)}_lb_l-\widehat{y}^{(i)}_l\sum_{j=1}^m K_{ij}\widehat{y}^{(j)}_lu_{l,j}+\delta^{(i)}\sum_{j=1}^{m} \sqrt{K_{jj}} \abs{u_{l,j}}\bigg)_i\bigg]\\
\text{s.t.} \quad & \gamma_l+1-\omega_{-l}\leq b_l \leq \gamma_l-1+\omega_{l}.
\end{aligned}
\end{equation}

\section{Computational Experiments} \label{sec_computational_results}
In this section, we discuss the performance of the deterministic models presented in Section \ref{sec_deterministic_SVM} and their robust counterparts of Section \ref{sec_robust_SVM} on a COVID-19 dataset. We start by describing the data collection process and their main characteristics (Section \ref{sec_comp_res_dataset_description}). Then, we present and analyze the results of the simulations on the basis of classical statistical indicators (Section \ref{sec_comp_res_model_validation}).

All the models were implemented in MATLAB (v. 2021b) and solved using CVX (v. 2.2, see \cite{GraBoy2008,cvx2014}) and MOSEK solver (v. 9.1.9, see \cite{mosek}). All computational experiments were run on an AMD EPYC 7302 Processor with 16-Core and 512 GB of RAM memory. Unless otherwise specified, a runtime limit of 48 hours (172800 seconds) is imposed. In models \eqref{linesearch_det_binary}, \eqref{linesearch_det_multiclass}, \eqref{linesearch_rob_binary} and \eqref{linesearch_rob_multiclass} we set the maximum number of subdivisions $N_{\max}$ equal to 100. 

\subsection{Dataset Description} \label{sec_comp_res_dataset_description}
Saliva samples, health records, and clinical data were acquired at \textit{IRCCS Fondazione Don Carlo Gnocchi ONLUS}, \textit{Santa Maria Nascente Hospital} in Milano (Italy), and \textit{Centro Spalenza Hospital} in Rovato (Italy), between April 2020 to July 2020. The COVID-19 diagnosis was conducted following the World Health Organization guidelines, declaring a positive case after the positive result of sequencing or Real-Time reverse-transcription Polymerase Chain Reaction assay of SARS-CoV-2 for nasopharyngeal swabs. The patients were considered COVID-19 negativized after two consecutive tests with negative results.

\begin{figure}[h!]
     \centering
  \begin{subfigure}[b]{0.49\textwidth}
         \centering
         \includegraphics[width=\textwidth]{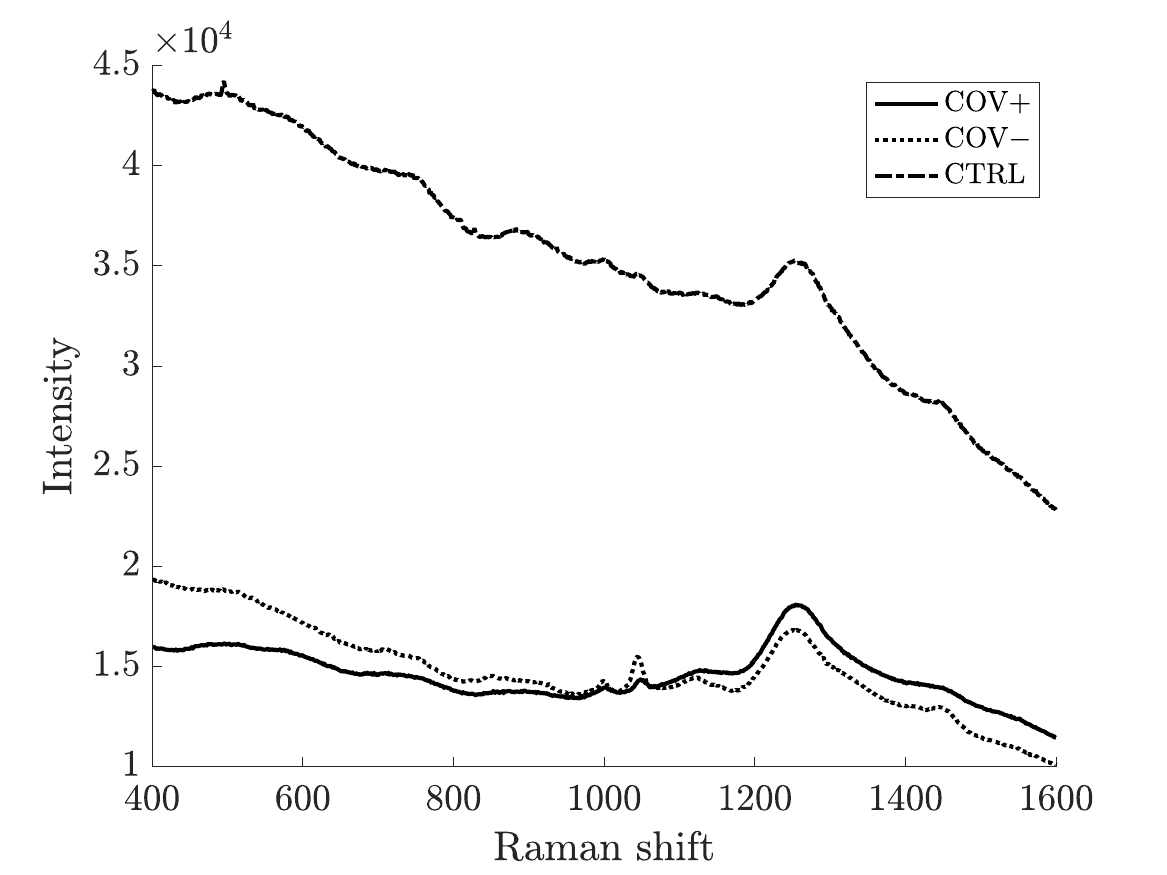}
         \caption{Average spectra of COV+, COV$-$ and CTRL patients.}
         \label{fig_spectra_cov}
     \end{subfigure}
     \begin{subfigure}[b]{0.49\textwidth}
         \centering
\includegraphics[width=\textwidth]{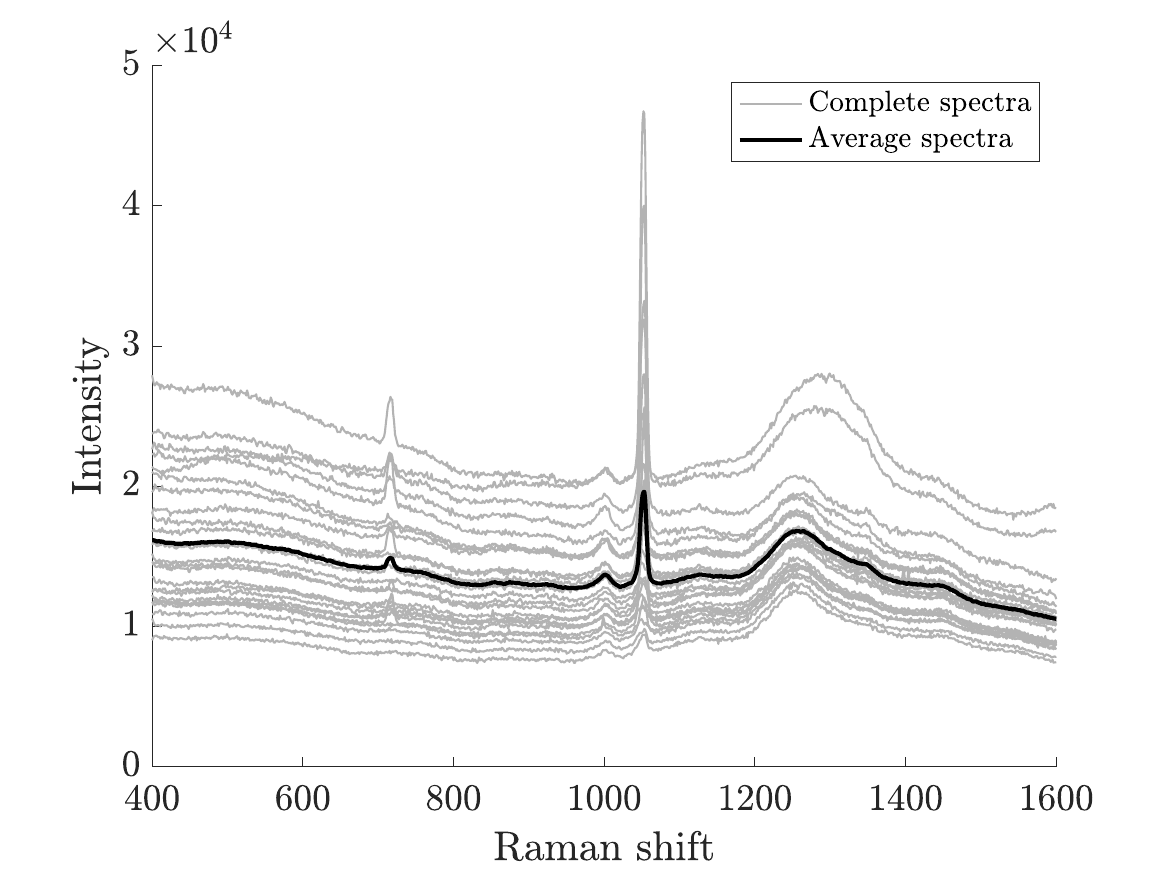}
         \caption{Complete and average spectra of a COV+ patient.}
         \label{fig_spectra_av}
     \end{subfigure}
          \caption{Examples of saliva Raman spectra in the dataset. On the left: average spectra of three patients (COV+, COV$-$ and CTRL). On the right: complete spectra and average spectra of a COV+ patient.}
        \label{fig_spectra}
\end{figure}

The total number of subjects involved in the study was 101, composed as follows: 30 patients affected by COVID-19 (COV+), 37 subjects negative to the SARS-CoV-2 test with an ascertained episode of COVID-19 (COV$-$), and 34 age and sex correlated healthy subjects (CTRL). More information regarding the acquisition protocol and the patients selections are available in \cite{cmess}. In Figure \ref{fig_spectra_cov} an example of three average Raman spectra is shown, each corresponding to a specific group of patients (COV+, COV$-$, CTRL).

Before applying the Machine Learning methods to our dataset, we perform the following cleaning and preprocessing steps (see \cite{raman_preproc, BERTAZIOLI2024108028}):

\begin{itemize}
    \item \textit{Outlier removal}: a spectrum is considered as an outlier when it encounters issues during the acquisition phase that lead to a low signal-to-noise ratio (see \cite{raman_preproc}). To ensure the integrity of the dataset, we excluded spectra that contained more than 10\% of values equal to zero or sequences of repeated values exceeding a saturation limit, defined as the maximum value that can be found within the single spectrum;
    \item \textit{Spike removal}: cosmic rays may sometimes negatively influence the measuring process, producing anomalous peaks in a spectrum. To remove them we exploited the Whitaker-Hayes algorithm (see \cite{spike}), where the series of subsequent differences in a spectrum is taken into account to highlight and remove peak anomalies;
    \item \textit{Realignment of the Raman shift axis}: the acquisition of Raman spectra may occur at different times and under varying conditions, potentially resulting in slight wavenumber shifts that misalign the Raman peaks. This issue can be addressed by interpolating the spectral data onto a common fixed grid of X-axis points;
    \item \textit{Removal of the background noise}: the acquisition of Raman spectra is negatively affected by background noise from fluorescence generated by molecules excited by the laser, which compromises the signal-to-noise ratio. Since this effect introduces wavenumber shifts that do not directly relate to the specific compound under investigation, it is usually recommended to remove them. Following recent Spectroscopy literature (see \cite{raman_preproc}) we employed polynomial fitting for this purpose (see \cite{polynomial_fitting}).
    \item \textit{Intensity normalisation}: normalisation techniques are employed to ensure consistent comparisons between Raman spectra collected under different conditions. Furthermore, since models \eqref{model_SVM_deterministic_binary}-\eqref{model_SVM_deterministic_multiclass} and their robust counterparts \eqref{model_SVM_robust_binary}-\eqref{model_SVM_robust_multiclass} are distance-based, imbalances in the magnitude of the features can lead to distorted classifiers. To mitigate these issues, in this study, we implemented the Standard Normal Variate Normalisation (see \cite{HanKamPei2011}), a widely used techniques in the field of spectroscopy.
    \item \textit{Principal Component Analysis}: for each of the 101 patients, approximately 25 salivary samples were acquired consisting in more than 900 components. Best practices often suggest avoiding the direct computation on high-dimensional raw data in ML applications (see \cite{HanKamPei2011}). For this reason, the most informative 15 features were extracted through a PCA method.
\end{itemize}

After cleaning and preprocessing the original dataset according to the previous steps, we also created a reduced dataset consisting of the average spectra for each patient (see Figure \ref{fig_spectra_av} for an illustrative example with a COV+ patient).

Summarizing, in this study we consider the following two datasets:
\begin{itemize}
    \item[a)] Complete COVID-19 dataset: $n=15$ features, $m=2409$ observations;
    \item[b)] Average COVID-19 dataset: $n=15$ features, $m=101$ observations.
\end{itemize}

\subsection{Model Validation} \label{sec_comp_res_model_validation}
The experimental setting is as follows. Each dataset was divided into training set and testing set through a \emph{Leave One Patient Out-Cross Validation} (LOPO-CV) approach. Specifically, in this study LOPO-CV can be seen as a 101-fold cross validation where all the patients except one were assigned to the training set. Once the classifier has been trained, its performance was tested on the unique patient in the testing set and the procedure was then repeated for all patients.

Regarding the kernel function $k(\cdot,\cdot)$, seven different alternatives were explored: homogeneous linear ($d=1$, $c=0$), homogeneous quadratic ($d=2$, $c=0$), homogeneous cubic ($d=3$, $c=0$); inhomogeneous linear, inhomogeneous quadratic, inhomogeneous cubic; Gaussian. Similarly to \cite{maggspin}, parameters $\alpha$ and $c$ (see Table \ref{tab_kernel}) were set as the maximum value of the standard deviation across features for the dataset under consideration.

Since the datasets are composed by three classes, we decided to consider four binary classification tasks (COV+ vs COV$-$; COV+ vs CTRL; COV$-$ vs CTRL; (COV+ $\cup$ COV $-$) vs CTRL) and a multiclass task (COV+ vs COV$-$ vs CTRL).

To measure the quality of the solution, we considered various statistical indicators, depending on the nature of the classification problem. Specifically, in the case of binary classification, let TP$_s$, TN$_s$, FP$_s$, FN$_s$ be the number of true positive, true negative, false positive and false negative, respectively, identified by the optimal classifier in fold $s$. Thus, for each fold $s=1,\ldots,101$, we computed the following indicators:
\begin{equation*}
\begin{aligned}
    & \text{Accuracy for fold } s := \text{A}_s = \displaystyle{\frac{\text{TP}_s+\text{TN}_s}{\text{TP}_s+\text{TN}_s+\text{FP}_s+\text{FN}_s}}
    \\
    & \text{Precision for fold } s := \text{P}_s = \displaystyle{\frac{\text{TP}_s}{\text{TP}_s+\text{FP}_s}}
    \\
    & \text{Sensitivity for fold } s := \text{SN}_s = \displaystyle{\frac{\text{TP}_s}{\text{TP}_s+\text{FN}_s}}
    \\
    & \text{Specificity for fold } s := \text{SP}_s = \displaystyle{\frac{\text{TN}_s}{\text{TN}_s+\text{FP}_s}}
    \\
    & \text{Matthews Correlation Coefficient for fold } s := \text{MCC}_s =
    \\
    & = \displaystyle{\frac{\text{TP}_s\cdot\text{TN}_s-\text{FP}_s\cdot\text{FN}_s}{\sqrt{(\text{TP}_s+\text{FP}_s)(\text{TP}_s+\text{FN}_s)(\text{TN}_s+\text{FP}_s)(\text{TN}_s+\text{FN}_s)}}}.
\end{aligned}
\end{equation*}

Finally, the results are averaged, leading to:
\begin{equation*}
\text{Accuracy} := \frac{1}{101} \sum_{s=1}^{101} \text{A}_s \qquad \text{Precision} := \frac{1}{101} \sum_{s=1}^{101} \text{P}_s \qquad \text{Sensitivity} := \frac{1}{101} \sum_{s=1}^{101} \text{SN}_s
\end{equation*}
\begin{equation*}
    \text{Specificity} := \frac{1}{101} \sum_{s=1}^{101} \text{SP}_s \qquad \text{MCC} := \frac{1}{101} \sum_{s=1}^{101} \text{MCC}_s.
\end{equation*}

Concerning the multiclass classification task, for each fold $s$ let C$_s^{\widehat{l},l}$ be the number of observations in class $l$ and classified in class $\widehat{l}$, with $l,\widehat{l}\in\{1,2,3\}=\{$COV+,COV$-$,CTRL\}. If $l=\widehat{l}$, then the observations are correctly classified, otherwise they are misclassified. Hence, for each fold $s=1,\ldots,101$ we computed:
\begin{equation*}
\begin{aligned}
    & \text{Accuracy for fold } s := \text{AC}_s = \displaystyle{\frac{\displaystyle\sum_{l=1}^3 \text{C}_s^{l,l}}{\displaystyle\sum_{\widehat{l},l=1}^3\text{C}^{\widehat{l},l}_s}}
    \\
    & \text{Sensitivity for class } l \text{ and fold } s := \text{SN}^l_s = \displaystyle{\frac{\text{C}_s^{l,l}}{\displaystyle\sum_{{\widehat{l}=1, \widehat{l}\ne l}}^3\text{C}^{\widehat{l},l}_s}}
    \\
    & \text{Specificity for class } l \text{ and fold } s := \text{SP}^l_s = \displaystyle{\frac{\displaystyle\sum_{{\widehat{l}=1, \widehat{l}\ne l}}^3 \displaystyle\sum_{{\widetilde{l}=1, \widetilde{l}\ne l}}^3\text{C}^{\widehat{l},\widetilde{l}}_s}{\displaystyle\sum_{\widehat{l}=1}^3 \displaystyle\sum_{{\widetilde{l}=1, \widetilde{l}\ne l}}^3\text{C}^{\widehat{l},\widetilde{l}}_s}}.
\end{aligned}
\end{equation*}

As in the binary case, we averaged the results as follows:
\begin{equation*}
\text{Accuracy} := \frac{1}{101} \sum_{s=1}^{101} \text{AC}_s \qquad \text{Sensitivity for class } l := \frac{1}{101} \sum_{s=1}^{101} \text{SN}^l_s
\end{equation*}
\begin{equation*}
        \text{Specificity for class } l := \frac{1}{101} \sum_{s=1}^{101} \text{SP}^l_s.
\end{equation*}

In the training phase, we explored two different approaches in treating hyperparameter $\nu$ in the objective function of models \eqref{model_SVM_deterministic_binary}, \eqref{model_SVM_deterministic_multiclass}, \eqref{model_SVM_robust_binary} and \eqref{model_SVM_robust_multiclass}: a grid search procedure and a Bayesian Optimization algorithm (see \cite{RAIAAN2024100470}).

In the first approach, five logarithmically spaced values between $10^{-3}$ and $10^0$ were considered (see \cite{FacMagPot2022,maggspin}), choosing the best one minimizing the training error. The results of the computations in terms of accuracy are reported in Table \ref{tab_accuracy_deterministic_gridsearch_ALLRESULTS} and specified on each dataset (complete COVID-19 dataset, see Table \ref{tab_accuracy_deterministic_gridsearch_COMPLETEDATASET}; average COVID-19 dataset, see Table \ref{tab_accuracy_deterministic_gridsearch_AVERAGEDATASET}).

\begin{table}[h]
    \begin{subtable}[h]{\textwidth}
        \centering
        \resizebox{\textwidth}{!}{
\begin{tabular}{ll *{7}l}\toprule
Classification task &  & \multicolumn{7}{c}{Kernel}\\
& & Hom. linear & Hom. quadratic & Hom. cubic & Inhom. linear & Inhom. quadratic & Inhom. cubic & Gaussian \\ \toprule
\multirow{2}{*}{COV+ vs COV$-$} & 
Accuracy (\%) & $74.27 \pm 34.79$ & $\underline{\mathbf{86.69 \pm 23.21}}$ & $69.93 \pm 27.80$ & $74.27 \pm 34.79$ & $86.43 \pm 22.91$ & $79.04 \pm 23.12$ & $81.04 \pm 28.25$ 
\\
& CPU time (s) & 2887&	3190&	25171&	3768&	3279&	7599&	3032
\\
\hline
\multirow{2}{*}{COV+ vs CTRL} & 
Accuracy (\%) & $\underline{\mathbf{80.96 \pm 32.40}}$&	$67.40 \pm 29.85$&	$67.31 \pm 30.99$&	$\underline{\mathbf{80.96 \pm 32.40}}$&	$69.90 \pm 34.12$&	$69.50 \pm 30.60$&	$78.74 \pm 33.96$
\\
& CPU time (s) & 2792&	6887&	24499&	3082&	3644&	14056&	2933
\\
\hline
\multirow{2}{*}{COV$-$ vs CTRL} & 
Accuracy (\%) & $82.38 \pm 29.73$&	$88.25 \pm 18.46$&	$86.18 \pm 19.41$&	$82.38 \pm 29.95$ &	$86.89 \pm 22.28$&	$86.25 \pm 20.08$&	$\underline{\mathbf{89.31 \pm 20.93}}$
\\
& CPU time (s) & 12449&	5858&	7352&	4310&	3727&	8127&	6618
\\
\hline
\multirow{2}{*}{(COV+ $\cup$ COV$-$) vs CTRL}& 
Accuracy (\%) & $\underline{\mathbf{82.32 \pm 29.51}}$	 & $77.12 \pm 29.19$ &	$74.40 \pm 27.69$ & $\underline{\mathbf{82.32 \pm 29.51}}$ & $80.56 \pm 28.59$ & $77.16 \pm 27.33$ & $66.34 \pm 47.49$
\\
& CPU time (s) & 2576 & 3612 &	11056 &	2651 &	4009 &	13771 &	2410
\\
\hline \hline
\multirow{2}{*}{COV+ vs COV$-$ vs CTRL}& 
Accuracy (\%) & $69.90 \pm 39.11$&	$71.55 \pm 31.89$&	$62.30 \pm 30.57$&	$69.90 \pm 39.11$&	$\underline{\mathbf{75.38 \pm 31.98}}$&	$68.69 \pm 29.22$&	$73.63 \pm 33.25$ 
\\
& CPU time (s) & 20189 &	20284 &	183159 & 20365 &	17461 &	271198&	12835
\\
\bottomrule
               \end{tabular}}
       \caption{Complete COVID-19 dataset.}
       \label{tab_accuracy_deterministic_gridsearch_COMPLETEDATASET}
    \end{subtable}
    \vfill \vspace*{0.25cm}
    \begin{subtable}[h]{\textwidth}
        \centering
        \resizebox{\textwidth}{!}{
\begin{tabular}{ll *{7}l}\toprule
Classification task &  & \multicolumn{7}{c}{Kernel}\\
& & Hom. linear & Hom. quadratic & Hom. cubic & Inhom. linear & Inhom. quadratic & Inhom. cubic & Gaussian \\ \toprule
\multirow{2}{*}{COV+ vs COV$-$} & 
Accuracy (\%) & $\underline{\mathbf{77.61 \pm 42.00}}$ &	$65.67\pm 47.84$ &	$65.67 \pm 47.84$ &	$\underline{\mathbf{77.61 \pm 42.00}}$ &	$67.16 \pm 47.32$	& $70.15 \pm 46.11$ & $44.78 \pm 50.10$
\\
& CPU time (s) & 15 &	15 & 16 & 15 & 15 & 17 & 15
\\
\hline
\multirow{2}{*}{COV+ vs CTRL} & 
Accuracy (\%) & $81.25 \pm 39.34$ &	$78.12\pm 41.67$ & $76.56\pm 42.70$ & $81.25 \pm 39.34$ & $75.00 \pm 43.64$ & $\underline{\mathbf{82.81 \pm 38.03}}$ & $46.88 \pm 50.30$
\\
& CPU time (s) & 15 & 15 & 15 & 15 & 14 & 15 & 15
\\
\hline
\multirow{2}{*}{COV$-$ vs CTRL} & 
Accuracy (\%) & $\underline{\mathbf{85.92 \pm 35.03}}$ & $80.28 \pm 40.07$ & $83.10 \pm 37.74$ & $\underline{\mathbf{85.92 \pm 35.03}}$ & $83.10 \pm 37.74$ & $78.87 \pm 41.11$ & $52.11 \pm 50.31$
\\
& CPU time (s) & 16  & 16 & 17 & 16 & 16 & 17 & 16
\\
\hline \hline
\multirow{2}{*}{COV+ vs COV$-$ vs CTRL}& 
Accuracy (\%) & $\underline{\mathbf{71.29 \pm 45.47}}$ & $70.30 \pm 45.92$ & $65.35 \pm 47.82$ & $\underline{\mathbf{71.29 \pm 45.47}}$ & $68.32 \pm 46.76$ & $69.31 \pm 46.35$ & $61.39 \pm 48.93$
\\
& CPU time (s) & 68 & 67 & 70 & 66 & 68 & 70 & 65
\\
\bottomrule
               \end{tabular}}
       \caption{Average COVID-19 dataset.}
       \label{tab_accuracy_deterministic_gridsearch_AVERAGEDATASET}
    \end{subtable}     
\caption{Out-of-sample accuracy and standard deviation obtained with the deterministic models \eqref{model_SVM_deterministic_binary}, \eqref{model_SVM_deterministic_multiclass} and a grid search procedure in tuning hyperparameter $\nu$. Best results are highlighted.} \label{tab_accuracy_deterministic_gridsearch_ALLRESULTS}
\end{table}

It can be noted that in both cases, the accuracy of the best classifier exceeds $71\%$, demonstrating that the considered SVM methodology is generally effective at correctly classifying the samples. However, when moving from dataset a) (complete COVID-19 dataset) to dataset b) (average COVID-19 dataset), the results worsen overall, except for the task COV+ vs CTRL, where the inhomogeneous cubic kernel achieves the highest accuracy in dataset b) at 82.81\% (compared to 80.96\% in dataset a) with both the homogeneous and inhomogeneous linear kernels). This general decline in accuracy reflects the reduced informative power of the average dataset compared to the original one. Additionally, in dataset b) the results tend to fluctuate more, showing larger standard deviations and indicating that the reduced data granularity makes it harder to distinguish between classes, especially for more complex tasks. On the other hand, in the reduced dataset CPU times are generally much lower, with most computations taking around 15–17 seconds for binary classification tasks and 65–70 seconds for the multiclass tasks. This shows significant computational efficiency compared to the complete dataset. Therefore, we can conclude that there exists a trade-off between accuracy and performing speed. The final user must balance these factors based on their priority: faster results with lower accuracy (average dataset) or more precise classifications with longer computation times (complete dataset).

A similar drop in performance is confirmed by other statistical indicators (see Table \ref{tab_other_deterministic_gridsearch_ALLRESULTS}), confirming that the reduced dataset leads to a decrease in model reliability. As far as it concerns the kernel functions, we notice that the Gaussian kernel tends to maximize sensitivity, reaching 100\% in several cases (see Table \ref{tab_other_deterministic_gridsearch_AVERAGEDATASET}), but it suffers from extremely poor specificity, making it unsuitable for balanced tasks and severely limiting its usefulness.
On the other hand, homogeneous and inhomogeneous polynomial kernels, especially linear and quadratic, generally achieve the best overall performance across multiple indicators, with particularly strong precision and MCC in most tasks, making them effective in both binary and multiclass problems.

\begin{table}[h]
    \begin{subtable}[h]{\textwidth}
        \centering
        \resizebox{\textwidth}{!}{
\begin{tabular}{ll *{7}l}\toprule
Classification task &  & \multicolumn{7}{c}{Kernel}\\
& & Hom. linear & Hom. quadratic & Hom. cubic & Inhom. linear & Inhom. quadratic & Inhom. cubic & Gaussian \\ \toprule
\multirow{4}{*}{COV+ vs COV$-$}& 
Precision (\%)& 76.13&	\underline{\textbf{87.21}}&	$-$& 76.13&	87.05&	82.19&	79.52
\\
& Sensitivity (\%)& 55.67&	55.11&	53.29&	55.67&	55.02&	53.28&	\textbf{\underline{58.17}}
\\
& Specificity (\%) & 44.33 & 44.89&	46.71&	44.33&	44.98&	\underline{\textbf{46.72}}&	41.83
\\
& MCC & 0.49&	\underline{\textbf{0.73}}&	0.40&	0.49&	\underline{\textbf{0.73}}&	0.58&	0.62
\\
\hline
\multirow{4}{*}{COV+ vs CTRL} & 
Precision (\%)& \underline{\textbf{80.73}}&	66.91&	66.04&	\underline{\textbf{80.73}}&	68.00&	69.29&	80.08
\\
& Sensitivity (\%)& 52.10&	54.09&	55.10&	52.10&	\underline{\textbf{56.03}}&	52.56&	50.46
\\
& Specificity (\%) & 47.90&	45.91&	44.90&	47.90&	43.97&	47.44&	\underline{\textbf{49.54}}
\\
& MCC & \underline{\textbf{0.62}}&	0.34&	0.33&	\underline{\textbf{0.62}}&	0.39&	0.38&	0.57
\\
\hline
\multirow{4}{*}{COV$-$ vs CTRL} &
Precision (\%)& 78.97&	87.23&	86.65&	78.97&	86.95&	86.89&	\underline{\textbf{87.47}}
\\
& Sensitivity (\%)& \underline{\textbf{49.05}} &	47.07&	45.51&	\underline{\textbf{49.05}}&	46.15&	45.75&	47.60
\\
& Specificity (\%) & 50.95&	52.93&	\underline{\textbf{54.49}}&	50.95&	53.85&	54.25&	52.40
\\
& MCC & 0.65&	0.76&	0.72&	0.65&	0.74&	0.72&	\underline{\textbf{0.78}}
\\
\hline
\multirow{4}{*}{(COV+ $\cup$ COV$-$) vs CTRL} &
Precision (\%)& 82.60 &	83.87 &	84.66 &	82.60 &\underline{\textbf{86.50}} & 85.84 & 67.40
\\
& Sensitivity (\%)& 72.02 & 71.67 & 68.39 & 72.02 & 70.47 & 69.16 & \underline{\textbf{100.00}}
\\
& Specificity (\%) & 27.98 & 28.33 & \underline{\textbf{{31.61}}} & 27.98 & 29.53 & 30.84 & 0.00
\\
& MCC & \underline{\textbf{0.60}} & 0.49 & 0.45 & \underline{\textbf{0.60}} & 0.56 & 0.5 & $-$
\\
\hline \hline
\multirow{6}{*}{COV+ vs COV$-$ vs CTRL}& 
Sensitivity COV+ (\%)& 59.84&	59.04&	47.93&	59.84&	\underline{\textbf{64.26}}&	55.02&	64.12
\\
& Sensitivity COV$-$ (\%)& 71.72&	85.97&	67.73&	71.72&	\underline{\textbf{86.43}}&	77.65&	80.39
\\
& Sensitivity CTRL (\%)& \underline{\textbf{77.58}}&	67.90&	69.30&	\underline{\textbf{77.58}}&	72.87&	71.21&	74.52
\\ \cmidrule{2-9}
& Specificity COV+ (\%) & 83.93&	85.93&	78.61&	83.93&	\underline{\textbf{87.00}}&	83.97&	84.47
\\
& Specificity COV$-$ (\%) & 84.99&	92.62&	87.34&	84.99&	\underline{\textbf{93.47}}&	87.73&	86.49
\\
& Specificity CTRL (\%) & 86.15&	80.95&	78.54&	86.15&	83.84&	82.25&	\underline{\textbf{89.75}}
\\
\bottomrule
\end{tabular}
}
       \caption{Complete COVID-19 dataset.}
       \label{tab_other_deterministic_gridsearch_COMPLETEDATASET}
    \end{subtable}
    \vfill \vspace*{0.25cm}
    \begin{subtable}[h]{\textwidth}
        \centering
        \resizebox{\textwidth}{!}{
\begin{tabular}{ll *{7}l}\toprule
Classification task &  & \multicolumn{7}{c}{Kernel}\\
& & Hom. linear & Hom. quadratic & Hom. cubic & Inhom. linear & Inhom. quadratic & Inhom. cubic & Gaussian \\ \toprule
\multirow{4}{*}{COV+ vs COV$-$}& 
Precision (\%)& \underline{\textbf{74.19}} &	65.22 & 62.64 & \underline{\textbf{74.19}} &	66.67 &	67.86 & 44.78
\\
& Sensitivity (\%)& 44.23 & 34.09 &	38.64 &	44.23 &	35.56 &	40.43 &	\textbf{\underline{100.00}}
\\
& Specificity (\%) & 55.77 & \underline{\textbf{65.91}} & 61.36 & 55.77 & 64.44 & 59.57 & 0.00
\\
& MCC & \textbf{\underline{0.55}} & 0.30 & 0.30 & \textbf{\underline{0.55}} & 0.33 & 0.39 &	$-$
\\
\hline
\multirow{4}{*}{COV+ vs CTRL} & 
Precision (\%)& 76.47 &	83.33 &	77.78 &	76.47 &	81.82 &	\underline{\textbf{91.30}} & 46.88
\\
& Sensitivity (\%)& 50.00 &	40.00 &	42.86 &	50.00 &	37.50&	39.62 &	\underline{\textbf{100.00}}
\\
& Specificity (\%) & 50.00 & 60.00 & 57.14 & 50.00 & \underline{\textbf{62.50}} & 60.38 & 0.00
\\
& MCC & 0.63	& 0.57 & 0.53 & 0.63 &	0.51	& \underline{\textbf{0.67}} & $-$
\\
\hline
\multirow{4}{*}{COV$-$ vs CTRL} & Precision (\%)& 86.49 &	89.66 &	87.88 &	86.49 &	\underline{\textbf{96.30}} &	82.35 &	52.11
\\
& Sensitivity (\%)& 52.46 &	45.61 &	49.15 &	52.46 &	44.07 &	50.00 &	\underline{\textbf{100.00}}
\\
& Specificity (\%) & 47.54 & 54.39 & 50.85 & 47.54 & \underline{\textbf{55.93}} & 50.00 & 0.00
\\
& MCC & \underline{\textbf{0.72}}	& 0.62	& 0.67 &	\underline{\textbf{0.72}} &	0.69	& 0.58	& $-$
\\
\hline \hline
\multirow{6}{*}{COV+ vs COV$-$ vs CTRL}& 
Sensitivity COV+ (\%)& 63.33 &	\underline{\textbf{73.33}} &	56.67 &	63.33 &	53.33 &	63.33 &	20.00
\\
& Sensitivity COV$-$ (\%)& \underline{\textbf{75.68}} &	64.86 &	67.57 & \underline{\textbf{75.68}}	& 67.57 &	70.27 &	\underline{\textbf{75.68}}
\\
& Sensitivity CTRL (\%)& 73.53 & 73.53 & 70.59 & 73.53 &	\underline{\textbf{82.35}} &	73.53 &	\underline{\textbf{82.35}}
\\ \cmidrule{2-9}
& Specificity COV+ (\%) & 88.87 &	78.85 &	78.73 &	88.87 &	84.36 &	81.54 &	\underline{\textbf{91.56}}
\\
& Specificity COV$-$ (\%) & 84.38 &	\underline{\textbf{89.06}} &	81.25 &	84.38 &	84.38&	81.25 &	70.31
\\
& Specificity CTRL (\%) & 83.61 &	87.88 &	88.50 &	83.61 &	83.91 &	\underline{\textbf{91.14}} &	79.64
\\
\bottomrule
               \end{tabular}}
       \caption{Average COVID-19 dataset.}
       \label{tab_other_deterministic_gridsearch_AVERAGEDATASET}
    \end{subtable}     
\caption{Detailed out-of-sample statistical indicators obtained with the deterministic models \eqref{model_SVM_deterministic_binary}, \eqref{model_SVM_deterministic_multiclass} and a grid search procedure in tuning hyperparameter $\nu$. Best results are highlighted.} \label{tab_other_deterministic_gridsearch_ALLRESULTS}
\end{table}

To further support the results of our proposal, in Table \ref{tab_comparison_deterministic_gridsearch_ALLDATASET} we conducted a comparison between the best results of Tables \ref{tab_accuracy_deterministic_gridsearch_ALLRESULTS}-\ref{tab_other_deterministic_gridsearch_ALLRESULTS} and the out-of-sample accuracy and precision provided by \emph{scikit-learn}, a popular ML library implemented in Python (see \cite{scikitlearn}). The comparison illustrates a clear advantage for the proposed approach across several tasks. For instance, in the case of distinguishing COV$-$ from CTRL in the average dataset, our model achieved a precision of 96.30\% compared to scikit-learn performance of 86.13\% (see Table \ref{tab_comparison_deterministic_gridsearch_AVERAGEDATASET}), indicating a superior ability to minimize false positives in this classification. Similarly, in the task of identifying (COV+ $\cup$ COV$-$) vs CTRL in the complete dataset, our method ensured a precision of 86.50\%, outperforming the 82.34\% of scikit-learn (see Table \ref{tab_comparison_deterministic_gridsearch_COMPLETEDATASET}). In terms of accuracy, our models consistently delivered better results in most tasks when applied to the complete dataset.

\begin{table}[h!]
    \begin{subtable}[h]{\textwidth}
        \centering
\resizebox{0.9\textwidth}{!}{
\begin{tabular}{ll *{2}l | *{2}l}\toprule
Classification task &  & \multicolumn{2}{c|}{Accuracy (\%)} & \multicolumn{2}{c}{Precision (\%)}\\
& & Table \ref{tab_accuracy_deterministic_gridsearch_COMPLETEDATASET} & Scikit-learn library & Table \ref{tab_other_deterministic_gridsearch_COMPLETEDATASET} & Scikit-learn library\\ \toprule
COV+ vs COV$-$ & & \underline{\textbf{86.69}} & 81.81 & \underline{\textbf{87.21}} & 81.82
\\ \hline
COV+ vs CTRL & & \underline{\textbf{80.96}} & 80.35 & \underline{\textbf{80.73}} & 80.35
\\ \hline
COV$-$ vs CTRL & & \underline{\textbf{89.31}} & 89.29 & 87.47 & \underline{\textbf{89.09}}
\\ \hline
(COV+ $\cup$ COV$-$) vs CTRL & & 82.32 & \underline{\textbf{84.36}}  & \underline{\textbf{86.50}} & 82.34
\\ \hline \hline
COV+ vs COV$-$ vs CTRL & & \underline{\textbf{75.38}} & 74.39  & $-$ & $-$
\\
\bottomrule
\end{tabular}
}
       \caption{Complete COVID-19 dataset.}
       \label{tab_comparison_deterministic_gridsearch_COMPLETEDATASET}
    \end{subtable}
    \vfill \vspace*{0.25cm}
    \begin{subtable}[h]{\textwidth}
        \centering
\resizebox{0.9\textwidth}{!}{
\begin{tabular}{ll *{2}l | *{2}l}\toprule
Classification task &  & \multicolumn{2}{c|}{Accuracy (\%)} & \multicolumn{2}{c}{Precision (\%)}\\
& & Table \ref{tab_accuracy_deterministic_gridsearch_AVERAGEDATASET} & Scikit-learn library & Table \ref{tab_other_deterministic_gridsearch_AVERAGEDATASET} & Scikit-learn library\\ \toprule
COV+ vs COV$-$ & & \underline{\textbf{77.61}} & 74.63 & 74.19 & \underline{\textbf{74.45}}
\\ \hline
COV+ vs CTRL & & 82.81 & \underline{\textbf{84.12}} & \underline{\textbf{91.30}} & 84.12
\\ \hline
COV$-$ vs CTRL & & 85.92 & \underline{\textbf{86.13}} & \underline{\textbf{96.30}} & 86.13
\\ \hline \hline
COV+ vs COV$-$ vs CTRL & & 71.29 & \underline{\textbf{74.04}}  & $-$ & $-$
\\
\bottomrule
\end{tabular}
}
       \caption{Average COVID-19 dataset.}
       \label{tab_comparison_deterministic_gridsearch_AVERAGEDATASET}
    \end{subtable}
\caption{Out-of-sample accuracy and precision comparison among the best results of Tables \ref{tab_accuracy_deterministic_gridsearch_ALLRESULTS}-\ref{tab_other_deterministic_gridsearch_ALLRESULTS} and simulations from the scikit-learn library (see \cite{scikitlearn}). Best results are highlighted.} \label{tab_comparison_deterministic_gridsearch_ALLDATASET}
\end{table}

As an alternative strategy for identifying the optimal value of the hyperparameter $\nu$, we adopted the Bayesian optimization approach. We utilized the \emph{bayesopt} library in Matlab, specifically configured to minimize the training error at each iteration. To ensure consistency with the grid search results, we restricted the search for $\nu$ to the interval [0,2]. The simulations were conducted using the best-performing kernel functions identified in the grid search procedure (see Table \ref{tab_accuracy_deterministic_gridsearch_ALLRESULTS}). For the robust SVM formulations (see Section \ref{sec_robust_SVM}), we employed a box-type uncertainty set ($p=\infty$ in definition \eqref{U_input_space}), assuming a constant uncertainty radius $\eta=\eta^{(i)}$ across all observations. The value of $\eta$ was tuned using Bayesian optimization over the interval $[10^{-7},10^{-1}]$ to identify the most robust configuration.

The results of the simulations are reported in Table \ref{tab_bayesianoptimization_ALLDATASET}. The robust SVM models provide slight improvements in statistical indicators like accuracy and sensitivity compared to the deterministic models across various classification tasks. However, regarding CPU time, Bayesian optimization techniques require nearly double the time compared to the grid search procedure for $\nu$ in the deterministic setting (see Table \ref{tab_accuracy_deterministic_gridsearch_ALLRESULTS}). This computational time is further increased in the robust framework due to the simultaneous tuning of both $\nu$ and $\eta$. In the multiclass classification task using the complete dataset, the Bayesian optimization method fails to provide results within the time limit of 48 hours (172800 seconds).

\begin{table}[h!]
\begin{subtable}[h]{\textwidth}
        \centering
\resizebox{0.65\textwidth}{!}{
\begin{tabular}{ll *{2}l |l}\toprule
Classification task &  Kernel & & Deterministic & Robust \\
\toprule
\multirow{6}{*}{COV+ vs COV$-$} &
\multirow{6}{*}{Hom. quadratic} &
Accuracy (\%) & $86.69 \pm 23.21$ & $\underline{\mathbf{87.00 \pm 22.08}}$
\\
& & Precision (\%) & \underline{\textbf{87.21}}  & 86.93
\\
& & Sensitivity (\%) & 55.11 & \underline{\textbf{55.52}}
\\
&  & Specificity (\%) & \underline{\textbf{44.89}} & 44.48
\\
& & MCC & 0.73 & \underline{\textbf{0.74}}
\\
& & CPU time (s) & 5568 & 20456
\\
\hline
\multirow{6}{*}{COV+ vs CTRL} & \multirow{6}{*}{Hom. linear} &
Accuracy (\%) & $80.96 \pm 32.40$ & $\underline{\mathbf{82.52 \pm 32.02}}$
\\
& & Precision (\%) & 80.50 & \underline{\textbf{82.66}}
\\
& & Sensitivity (\%) & \underline{\textbf{52.10}} & 51.70
\\
&  & Specificity (\%) & 47.90 & \underline{\textbf{48.30}}
\\
& & MCC & 0.62 & \underline{\textbf{0.65}}
\\
& & CPU time (s) & 5351 & 13485
\\
\hline
\multirow{6}{*}{COV$-$ vs CTRL} & 
\multirow{6}{*}{Gaussian} &
Accuracy (\%) & $\underline{\mathbf{79.63 \pm 28.61}}$  & $79.54 \pm 29.24$
\\
& & Precision (\%) & \underline{\textbf{92.45}} & 92.12
\\
& & Sensitivity (\%) & \underline{\textbf{37.05}} & 36.43
\\
& & Specificity (\%) & 62.95 & \underline{\textbf{63.57}}
\\
& & MCC & \underline{\textbf{0.61}} & \underline{\textbf{0.61}}
\\
& & CPU time (s) & 13143 & 26616
\\
\hline
\multirow{6}{*}{(COV+ $\cup$ COV$-$) vs CTRL} & 
\multirow{6}{*}{Hom. linear} &
Accuracy (\%) & $\underline{\mathbf{82.35 \pm 29.23}}$  & $82.20 \pm 29.60$
\\
& & Precision (\%) & \underline{\textbf{86.34}} & 86.05
\\
& & Sensitivity (\%) & 71.88 & \underline{\textbf{72.13}}
\\
& & Specificity (\%) & \underline{\textbf{28.12}} & 27.87
\\
& & MCC & \underline{\textbf{0.60}} & 0.59
\\
& & CPU time (s) & 13633 & 87744
\\
\hline \hline
\multirow{8}{*}{COV+ vs COV$-$ vs CTRL} &
\multirow{8}{*}{Inhom. quadratic} 
& Accuracy (\%) & $75.22 \pm 31.66$ & $-$
\\
& & Sensitivity COV+ (\%) & 62.25 & $-$
\\
& & Sensitivity COV$-$ (\%) & 87.00 & $-$
\\
& & Sensitivity CTRL (\%) & 73.76  & $-$
\\
& & Specificity COV+ (\%) & 87.88  & $-$
\\
& & Specificity COV$-$ (\%) & 92.75 & $-$
\\
& & Specificity CTRL (\%) & 83.45 & $-$
\\
& & CPU time (s) & 145923 & 172800*
\\
\bottomrule
\end{tabular}
}
       \caption{Complete COVID-19 dataset.}
       \label{tab_bayesianoptimization_COMPLETEDATASET}
    \end{subtable}
    \vfill \vspace*{0.25cm}
    \begin{subtable}[h]{\textwidth}
        \centering
\resizebox{0.65\textwidth}{!}{
\begin{tabular}{ll *{2}l |l}\toprule
Classification task &  Kernel & & Deterministic & Robust \\
\toprule
\multirow{6}{*}{COV+ vs COV$-$} &
\multirow{6}{*}{Hom. linear} &
Accuracy (\%) & $74.63 \pm 43.84$ & $\underline{\mathbf{76.12 \pm 42.96}}$
\\
& & Precision (\%) & 69.70 & \underline{\textbf{71.88}}
\\
& & Sensitivity (\%) & \underline{\textbf{46.00}} & 45.10
\\
&  & Specificity (\%) & 54.00 & \underline{\textbf{54.90}}
\\
& & MCC & 0.49 & \underline{\textbf{0.52}}
\\
& & CPU time (s) & 710 & 755
\\
\hline
\multirow{6}{*}{COV+ vs CTRL} & \multirow{6}{*}{Hom. cubic} &
Accuracy (\%) & $\underline{\mathbf{81.25 \pm 39.34}}$ & $59.38 \pm 49.50$
\\
& & Precision (\%) & \underline{\textbf{87.50}} & 56.90
\\
& & Sensitivity (\%) & 40.38 & \underline{\textbf{86.84}}
\\
&  & Specificity (\%) & \underline{\textbf{59.62}} & 13.16
\\
& & MCC & \underline{\textbf{1.03}} & 0.23
\\
& & CPU time (s) & 757 & 702
\\
\hline
\multirow{6}{*}{COV$-$ vs CTRL} & 
\multirow{6}{*}{Hom. linear} &
Accuracy (\%) & $\underline{\mathbf{85.92 \pm 35.03}}$  & $83.10 \pm 37.74$
\\
& & Precision (\%) & 86.49 & \underline{\textbf{89.29}}
\\
& & Sensitivity (\%) & \underline{\textbf{52.46}} & 42.37
\\
& & Specificity (\%) & 47.54 & \underline{\textbf{57.63}}
\\
& & MCC & \underline{\textbf{1.12}} & 1.07
\\
& & CPU time (s) & 736 & 807
\\
\hline \hline
\multirow{8}{*}{COV+ vs COV$-$ vs CTRL} &
\multirow{8}{*}{Hom. linear} 
& Accuracy (\%) & $\underline{\mathbf{71.29 \pm 45.47}}$ & $68.32 \pm 46.76$
\\
& & Sensitivity COV+ (\%) & \underline{\textbf{63.33}} & 53.33
\\
& & Sensitivity COV$-$ (\%) & \underline{\textbf{72.97}} & \underline{\textbf{72.97}}
\\
& & Sensitivity CTRL (\%) & \underline{\textbf{76.47}}  & \underline{\textbf{76.47}}
\\
& & Specificity COV+ (\%) & 85.82  & \underline{\textbf{85.94}}
\\
& & Specificity COV$-$ (\%) & 82.21 & \underline{\textbf{85.94}}
\\
& & Specificity CTRL (\%) & \underline{\textbf{88.19}} & 81.27
\\
& & CPU time (s) & 5504 & 6677
\\
\bottomrule
\end{tabular}
}
       \caption{Average COVID-19 dataset.}
       \label{tab_bayesianoptimization_AVERAGEDATASET}
    \end{subtable}
\caption{Detailed results of average out-of-sample statistical measures obtained with the deterministic and robust models. Bayesian optimization techniques was used in tuning hyperparameter $\nu$. The asterisk indicates that the time limit has been reached. Best results are highlighted.} \label{tab_bayesianoptimization_ALLDATASET}
\end{table}

Finally, in Table \ref{tab_comparison_det_grid_bayesian_ALLDATASET} we summarize and compare the best results in terms of accuracy of this study.

\begin{table}[h!]
\begin{subtable}[h]{\textwidth}
        \centering
\resizebox{\textwidth}{!}{
\begin{tabular}{lcccc}\toprule
Classification task & Scikit-learn library & \multicolumn{2}{c}{Deterministic model} & Robust model\\
& & Grid-search & Bayesian Optimization & Bayesian Optimization\\
\toprule
COV+ vs COV$-$ & 81.81 & 86.69 & 86.69 & \underline{\textbf{87.00}}
\\ \hline
COV+ vs CTRL & 80.35 & 80.86 & 80.96 & \underline{\textbf{82.52}}
\\ \hline
COV$-$ vs CTRL & 89.29 & \underline{\textbf{89.31}} & 79.63 & 79.54
\\ \hline
(COV+ $\cup$ COV$-$) vs CTRL & \underline{\textbf{84.36}} & 82.32 & 82.35 & 82.20
\\ \hline \hline
COV+ vs COV$-$ vs CTRL & 74.39 & \underline{\textbf{75.38}} & 75.22 & $-$
\\
\bottomrule
\end{tabular}
}
       \caption{Complete COVID-19 dataset.}
       \label{tab_comparison_det_grid_bayesian_COMPLETEDATASET}
    \end{subtable}
    \vfill \vspace*{0.25cm}
\begin{subtable}[h]{\textwidth}
        \centering
\resizebox{\textwidth}{!}{
\begin{tabular}{lcccc}\toprule
Classification task & Scikit-learn library & \multicolumn{2}{c}{Deterministic model} & Robust model\\
& & Grid-search & Bayesian Optimization & Bayesian Optimization\\
\toprule
COV+ vs COV$-$ & 74.63 & \underline{\textbf{77.61}} & 74.63 & 76.12
\\ \hline
COV+ vs CTRL & \underline{\textbf{84.12}} & 82.81 & 81.25 & 59.38
\\ \hline
COV$-$ vs CTRL & \underline{\textbf{86.13}} & 85.92 & 85.92 & 83.10
\\ \hline \hline
COV+ vs COV$-$ vs CTRL & \underline{\textbf{74.04}} & 71.29 & 71.29 & 68.32
\\
\bottomrule
\end{tabular}
}
       \caption{Average COVID-19 dataset.}
       \label{tab_comparison_det_grid_bayesian_AVERAGEDATASET}
    \end{subtable}
\caption{Out-of-sample accuracy comparison among the best results of Tables \ref{tab_accuracy_deterministic_gridsearch_ALLRESULTS}, \ref{tab_other_deterministic_gridsearch_ALLRESULTS}, \ref{tab_bayesianoptimization_ALLDATASET} and simulations from the scikit-learn library (see \cite{scikitlearn}). Best results are highlighted.} \label{tab_comparison_det_grid_bayesian_ALLDATASET}
\end{table}

In the complete dataset (see Table \ref{tab_comparison_det_grid_bayesian_COMPLETEDATASET}), the robust models consistently outperformed the deterministic models in the COV+ vs COV$-$ and COV+ vs CTRL tasks. In all the other tasks the best results are in favour of deterministic approaches. On the other hand, in the average dataset (see Table \ref{tab_comparison_det_grid_bayesian_AVERAGEDATASET}), the results exhibit a different trend. In three out of four cases, the scikit-learn library outperforms the other methods. However, as pointed out, the informative power of this dataset is reduced.
These results highlight the trade-offs associated with using robust models and Bayesian optimization techniques. While they can improve accuracy in certain classification tasks, there are situations where deterministic approaches perform better.

\clearpage
\newpage
\section{Conclusions} \label{sec_conclusions}

In this paper, we presented data-driven optimization models to support medical decision-making in diagnosing COVID-19 through a combination of Raman Spectroscopy and Machine Learning methods. Specifically, we applied a novel approach proposed in \cite{maggspin} to handle Support Vector Machines with nonlinear decision boundaries. To account for uncertainties in saliva Raman spectra, we formulated robust optimization classifiers for both binary and multiclass classification tasks. We conducted numerical experiments based on real-world data on COVID-19 diagnosis provided by Italian hospitals. To evaluate the effectiveness of the proposal, we compared the results with a state-of-the-art classifier in Machine Learning applications. The experiments highlight a trade-off between computational efficiency and classification accuracy. Additionally, we explored two methods for tuning the hyperparameters of the models: a grid-search procedure and a Bayesian Optimization algorithm. The combination of Bayesian optimization and robust Support Vector Machine models led to small but consistent improvements in accuracy.

Overall, this work highlights the potential of Machine Learning techniques, especially robust Support Vector Machines, in improving disease detection through Raman spectroscopy with noisy and limited data. From a methodological perspective, future research could extend this approach to address uncertainties in the labels of spectral data, enhancing the model's generalization capabilities. Additionally, the potential of combining Machine Learning and Raman Spectroscopy can be explored for designing a rapid, cost-effective, and non-invasive diagnostic tool. Indeed, with recent advancements in spectroscopy and the development of portable Raman Spectroscopy devices, a point-of-care system could be established for use in any setting, without requiring specific human expertise. Finally, given the critical nature of healthcare applications, incorporating explainability in Machine Learning methods is essential to make the proposed models more transparent.

\section*{Founding}

This work has been supported by Fondazione Regionale per
la Ricerca Biomedica, project CORSAI ID 383 - JTC 2021 ERA PerMed, GA 779282 - CUP: H45E22000030006.

\section*{Acknowledgements}

MP was partially supported by the MUR under the grant “Dipartimenti di Eccellenza 2023-2027" of the Department of Informatics, Systems and Communication of the University of Milano-Bicocca, Italy

FM and AS  have been supported by ``ULTRA OPTYMAL - Urban Logistics and sustainable TRAnsportation: OPtimization under uncertainTY and MAchine Learning'', a PRIN2020 project funded by the Italian University and Research Ministry (grant number 20207C8T9M).

\bibliographystyle{elsarticle-num}

\bibliographystyle{spmpsci}

\bibliography{main}

\end{document}